\documentclass[conference, 10pt]{IEEEtran}
\usepackage{amsthm}

\usepackage{enumitem}
\newcounter{problem}
\newcounter{subproblem}[problem]

\usepackage{hyperref}
\usepackage{cleveref}

\usepackage{mathtools}

\usepackage{color}
\usepackage{xcolor}
\usepackage{colortbl}
\definecolor{purple}{RGB}{139, 0, 139}
\usepackage{manfnt}
\usepackage{cleveref}
\usepackage{url}

\usepackage{caption}
\usepackage{subcaption}

\newif\iftodo   
\todotrue
\newif\iftodoshort  
\todoshortfalse
\ifCLASSINFOpdf
  \usepackage[pdftex]{graphicx}
   \DeclareGraphicsExtensions{.pdf,.jpeg,.png,.tiff}
\else
   \usepackage[dvips]{graphicx}
   \DeclareGraphicsExtensions{.eps,.ps}
\fi
\usepackage{amssymb,amsmath}
\usepackage[mathscr]{euscript}
\usepackage{bbm}
\usepackage{dsfont}

\usepackage{algorithmic}
\usepackage[ruled,vlined,commentsnumbered]{algorithm2e}

\graphicspath{ {./fig/} }
\newcommand{\Rmnum}[1]{\uppercase\expandafter{\romannumeral #1}}
\newcommand{\rmnum}[1]{\lowercase\expandafter{\romannumeral #1}}

\usepackage[utf8]{inputenc}

\usepackage{hyperref}
\usepackage{amsthm}
\usepackage{amssymb,amsmath}
\usepackage[mathscr]{euscript}

\newcommand{\cosl}[1]{}
\newcommand{\resl}[1]{}

\usepackage[draft,footnote,nomargin]{fixme}
\newcommand{\fnql}[1]{}
\newcommand{\fnsv}[1]{}

\begin{document}
%
\title{A Statistical Approach for RF Exposure Compliance Boundary Assessment in Massive MIMO Systems}
\author{
\IEEEauthorblockN{Paolo Baracca, Andreas Weber, Thorsten Wild}
\IEEEauthorblockA{Nokia Bell Labs, Stuttgart, Germany\\
Email: \{paolo.baracca, andreas.weber, thorsten.wild\}@nokia-bell-labs.com}
\and
\IEEEauthorblockN{Christophe Grangeat}
\IEEEauthorblockA{Nokia Mobile Networks, Paris-Saclay, France\\
Email: christophe.grangeat@nokia.com}
}

\maketitle
\begin{abstract}
Massive multiple-input multiple-output (MIMO) is a fundamental enabler to provide high data throughput in next generation cellular networks. By equipping the base stations (BSs) with tens or hundreds of antenna elements, narrow and high gain beams can be used to spatially multiplex several user equipment (UE) devices. While increasing the achievable performance, focusing the transmit power into specific UE directions also poses new issues when performing the radio frequency (RF) exposure assessment. In fact, the spatial distribution of the actual BS transmit power strongly depends on the deployment scenario and on the position of the UEs. Traditional methods for assessing the RF exposure compliance boundaries around BS sites are generally based on maximum transmit power and static beams. In massive MIMO systems, these approaches tend to be very conservative, in particular when time averaging is properly considered. In this work, we propose to leverage the three dimensional spatial channel model standardized by the Third Generation Partnership Project in order to assess reasonably foreseeable compliance boundaries of massive MIMO BSs. The analysis is performed by considering BSs fully loaded and different configurations of active UEs per cell. Numerical results show that the statistical approach developed in this paper allows reducing to nearly half the compliance distance when compared to the traditional method.
\end{abstract}

\section{Introduction}

Massive multiple-input multiple-output (MIMO) is an important enabler to cope with the ever-increasing demand for data throughput in the fifth generation (5G) of cellular systems \cite{larsson_cm14}. Equipping base stations (BSs) with a high number of antennas allows to strongly increase the cell spectral efficiency, mainly thanks to two complementary techniques: beamforming and spatial multiplexing. By applying beamforming, the BS focuses the transmit energy toward the specific user equipment (UE) location, thus strongly increasing the receive signal power. With spatial multiplexing, multiple streams are sent by the BS toward several active UEs, which are separated by using different beamformers. Because of the huge benefits promised by massive MIMO, many papers and works have been carried out in the last years both in the academia and in the industry. Thanks to that, BSs equipped with tens or hundreds of antennas are expected to be deployed in the next few years, as some basic techniques like full-dimension MIMO have already been standardized in Release 13 of the Long Term Evolution (LTE) standard \cite{nam_cm12}. In parallel to the methods studied to improve performance, an important topic related to wireless communications at large is the radio frequency (RF) exposure of humans \cite{sambo_st15}. Few years ago the RF electromagnetic field (EMF) has been classified by the World Health Organization as “possibly carcinogenic to humans” (Group 2B) \cite{emf_2b}. Indeed, there are regulations specifying constraints on two metrics related to the EMF: the specific absorption rate (SAR), usually considered in the near-field region, e.g., for the EMF radiated by the UEs, and the power density, usually considered in the far-field region, e.g., for the EMF radiated by the BSs \cite{vallauri_map15}. The methods for assessing the RF exposure from BSs, which are specified by the International Electrotechnical Commission (IEC) \cite{iec_fdis62232_new_grangeat}, address the exposure limits defined, for example, by the International Commission on Non-Ionizing Radiation Protection \cite{icnirp_guidelines_grangeat}, and have been adopted in many countries and regions (including Europe), each detailing its own specific regulations.

A traditional approach exploited by operators to meet the EMF regulations at the BS sites is to design a compliance boundary, also known in the literature as exclusion zone, around the site and ensure no access to this area to the general public \cite{blanch_tap12}. This compliance boundary was traditionally designed in a conservative approach by considering the maximum theoretical transmit power in all the directions \cite{iec_fdis62232_new_grangeat}.

In fact, massive MIMO, while strongly increasing system performance, it allows focusing energy with sharp high gain beams into the specific UE directions. Therefore, designing the compliance boundary by using the maximum transmit power in all the directions would result in not realistic large areas \cite{degirmenci_tec16}, making problematic for the operators to deploy massive MIMO BSs on sites with pre-existing BSs, like the Global System for Mobile Communications (GSM), Universal Mobile Telecommunications System (UMTS) or LTE ones. On the other hand, the assessment of RF exposure is averaged over several minutes before comparing it to the allowed limits \cite{iec_fdis62232_new_grangeat}, and this is a very long period with respect to methods like beamforming update and UE scheduling, which happen at the BS every few milliseconds or even less. Because of that, very recently it has been recognized that the traditional approach for designing the compliance boundary might be over-conservative for at least two main reasons. First, BSs are not always 100\% loaded, so they do not transmit at full power in each millisecond, and then, when multiple UEs are served by a BS, even with spatial multiplexing, the power is split among different directions.

Therefore, statistical approaches for designing the compliance boundary at the BS have been introduced in \cite{iec_fdis62232_new_grangeat} to better assess the actual transmit power in more realistic scenarios: by taking into account massive MIMO BS operations, the compliance boundary turns out to be actually smaller when compared to the one computed with the conservative traditional method.

The implementation principles of statistical approaches are currently being investigated by international standardization bodies like the IEC to understand their benefits and limits for RF exposure assessment. In fact, most of the works on RF exposure with MIMO BSs consider the traditional approach. For instance, \cite{perentos_tap12} compares different summation schemes to estimate the RF exposure levels in MIMO systems, whereas \cite{thors_bio14} evaluates different assessment methods by considering MIMO BSs in a multi-band network. Just limited literature considers the statistical approaches. Only very recently \cite{thors_acc17} proposed a model for designing the compliance boundary of a massive MIMO system by considering the 95th percentile of the transmit power distribution as a reference. However, that study makes some simplified assumptions regarding both the beamforming design and the traffic conditions in order to derive an analytic expression of the actual transmit power. More work is needed in this direction to provide accurate assessment of practical RF exposure conditions: not dealing properly with the EMF constraints might become the biggest obstacle in preventing pervasive deployment of massive MIMO BSs in 5G and beyond cellular networks.

In this work, we consider a cellular network with massive MIMO BSs and provide extensive system level simulations to understand the benefits of statistical approaches in assessing the compliance boundary around the BSs. More specifically, we consider deployment scenarios and channel models standardized by the Third Generation Partnership Project (3GPP) and evaluate how the power is focused in a practical system when realistic assumptions regarding UE distribution and traffic models are taken into account. For each scenario, we derive the cumulative distribution function (CDF) of the power transmitted by the BSs. Then, we compute the compliance boundary based on a given percentile of the transmit power. When looking at the 95th or 99th percentile, numerical results show that the radius of the compliance boundary with this statistical approach can be nearly half of the one computed with the conservative traditional approach.

\section{Compliance Boundary Design based on Statistical Approaches}

Various methods have been defined in \cite{iec_fdis62232_new_grangeat} for the calculation of the BS compliance boundary: for the sake of simplicity, we implement here the far-field calculation method. In the far-field, the electromagnetic waves propagate at the speed of light and electric and magnetic fields are mutually perpendicular, i.e., only one of the two must be evaluated based on the power density.

Let us consider a spherical coordinate system with origin where the massive MIMO BS is located and denote with $G(\theta,\phi,t)$ the beamforming gain provided by the BS at the azimuth angle $\theta$, elevation angle $\phi$ and time $t$, where $(\theta,\phi)=(0^{\circ},0^{\circ})$ defines the direction perpendicular to the BS array. The power density generated by the BS can be written as \cite[(4)]{sambo_st15}
\begin{equation}
S(\theta,\phi,t) = \frac{E(\theta,\phi,t)^2}{Z_0} = \frac{P_{TX}G(\theta,\phi,t)}{4 \pi r^2}\,,
\label{eq_pwdens}
\end{equation}
where $E(\theta,\phi,t)$ is the electric field, $Z_0=120\pi$ the free space impedance, $P_{\rm TX}$ the maximum BS transmit power, and $r$ the distance from the BS.

As guidelines \cite{icnirp_guidelines_grangeat} recommend time averaging of RF exposure over 6 minutes, we indicate here the 6 minute average beamforming gain at direction $(\theta,\phi)$ as $G_{\rm A}(\theta,\phi)$, which is computed by averaging the instantaneous power density $S(\theta,\phi,t)$ and applying simple basic manipulations in (\ref{eq_pwdens}). Note that $G_{\rm A}(\theta,\phi)$ takes into account also the time slots where BSs do not transmit, for instance either because there is no data to be sent or because these slots are allocated to uplink (UL) transmissions in a time division duplex (TDD) setup. Although computed with an averaging over a rather long time window, $G_{\rm A}(\theta,\phi)$ is anyhow a random variable whose distribution (numerically evaluated in Section \ref{perf_sect}) depends on many system parameters, but mainly on the traffic conditions in the cells.
 
With the statistical approaches, the compliance boundary is computed by looking at the space around the BS such that the power density meets the regulator constraints within a pre-defined percentile $p$. By indicating with $E_{MAX}$ the electric field reference level as defined in \cite{icnirp_guidelines_grangeat} at the system carrier frequency, we can write from (\ref{eq_pwdens}) the distance $r_{p}(\theta,\phi)$ at which the EMF constraint is statistically met as
\begin{equation}
r_{p}(\theta,\phi) = \frac{1}{E_{MAX}} \sqrt{ \frac{ P_{TX} G_{{\rm A},p}(\theta,\phi) Z_0 }{4\pi} } \,,
\end{equation}
where $G_{{\rm A},p}(\theta,\phi)$ denotes the $p$-percentile of $G_{\rm A}(\theta,\phi)$.

In this work, we are interested in comparing the dimension of the compliance boundary when the statistical approaches are enforced and therefore we simply focus on the distance at which the maximum power density is observed, i.e.,
\begin{equation}
r_{{\rm CB},p} = \max_{\theta,\phi} \, r_{p}(\theta,\phi)\,.
\label{eq_rez}
\end{equation}
For the sake of clarity, we refer in the rest of paper to the distance defined in (\ref{eq_rez}) as {\it compliance distance}.

In order to provide a more accurate shape of the compliance boundary, more sophisticated methods such as the synthetic model method \cite{altman_tec02} or others defined in \cite{iec_fdis62232_new_grangeat} can also be implemented.

\section{System Setup}

In this work we assume the three dimensional (3D) spatial channel model proposed by 3GPP for the BS-UE link \cite{3gpp_tr36873_grangeat}. In this framework, we consider a hexagonal BS deployment with seven sites, three sectors per site and wraparound. We focus on two scenarios: urban macro (UMa) and urban micro (UMi). In UMa, BSs are deployed on top of buildings at a height of 25 m with an inter-site distance (ISD) of 500 m, whereas in UMi BSs are deployed at a smaller height of 10 m, but more densely with an ISD of 200 m.
The system works at a carrier frequency of 2 GHz and BSs transmit with power $P_{\rm TX}=49/44$ dBm in UMa/UMi on a system bandwidth of 20 MHz.
In both scenarios, each BS is equipped with 64 co-polarized antennas, arranged in a 8x8 array with a mechanical downtilt of $0^{\circ}$ and an antenna spacing of $0.5\lambda$, where $\lambda$ is the wavelength of the carrier frequency. Each antenna element is assumed to have a parabolic radiation pattern with 8 dBi gain, a half-power beamwidth of $65^{\circ}$ and a front-to-back ratio attenuation of 30 dB.
Note that, at the considered carrier frequency, the maximum electric field allowed by regulators is $E_{MAX}=61$ V/m \cite{icnirp_guidelines_grangeat}.

Single-antenna UEs are uniformly distributed in the coverage area of each BS: only 20\% of the active UEs are outdoor, whereas the remaining 80\% of the UEs are indoor in buildings whose height is uniformly distributed between 4 and 8 floors. We target a low-mobility scenario where the UE speed is 3 km/h.

While moving away from the traditional method, the statistical approaches need anyhow to focus on scenarios where the RF exposure can be a realistic issue. Therefore, we consider the case where BSs are fully loaded and assume the full-buffer traffic model by simulating a sequence of independent UE drops and denoting with $D$ the drop duration. We consider $K$ active UEs per cell, which are served in each time slot on the whole band with equal power assigned to each UE. The UEs are spatially separated by using eigen-beamforming \cite{zhou_it03}, where the beamformer for a specific UE a) is designed based on the covariance matrix of the channel and b) focuses the energy toward the strongest eigen-direction.

In Section \ref{perf_sect} we will investigate in more detail the impact of these two parameters $D$ and $K$ on the dimension of the compliance boundary. However, it is clear that, for a given direction $(\theta,\phi)$, the variance of the average beamforming gain $G_{\rm A}(\theta,\phi)$ will decrease when either $K$ increases or $D$ decreases. Indeed, both higher values of $K$ and smaller values of $D$ are related to a system with more UEs served by the BSs over the fixed time window of 6 minutes, which, in turn, means a higher space averaging.

\section{System Performance Evaluation}
\label{perf_sect}

\begin{figure}[t!]
\centering
\includegraphics[width=1\hsize]{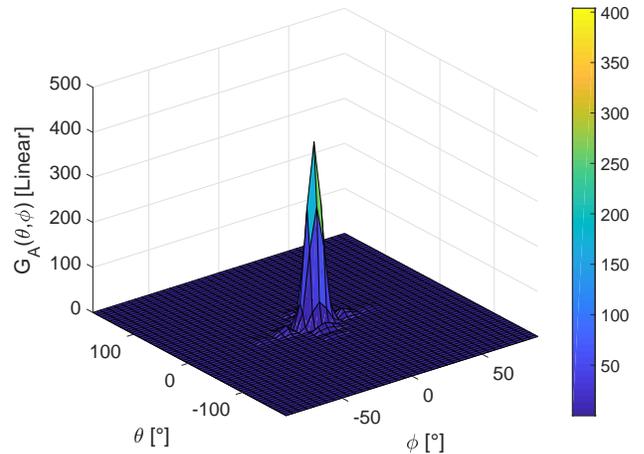}\\
\caption{Realization of $G_{\rm A}(\theta,\phi)$ in an artificial case where the BS employs a beamformer to focus the whole energy toward the direction $(\theta,\phi)=(0^{\circ},0^{\circ})$ perpendicular to the BS array.}
\label{fig_rp_art}
\end{figure}

In Fig. \ref{fig_rp_art} we plot the radiation pattern, i.e., the realization of $G_{\rm A}(\theta,\phi)$, of a BS using for the whole time window of 6 minutes only one beamformer which focuses the energy toward the direction perpendicular to the BS array $(\theta,\phi)=(0^{\circ},0^{\circ})$. While this full power static scenario has been artificially built, it represents a typical situation used to define the compliance boundary with the traditional method. The radiation pattern turns out here to be made by only one very sharp and high gain beam: in the direction of maximum we have a beamforming gain in linear scale of about 400, which corresponds to a gain of about 26 dBi in logarithmic scale. This high value is expected here, as BSs are equipped with 64 antennas, thus providing an array gain of about 18 dBi, which must be added up to the antenna element gain of 8 dBi. As a consequence, in this setup representing the traditional method to compute the compliance boundary, the compliance distance (\ref{eq_rez}) for the two scenarios under study is $r_{{\rm CB,UMa}}^{\rm (trad)}=16.1$ m and $r_{{\rm CB,UMi}}^{\rm (trad)}=9$ m. Note that, these two distances have been computed by assuming a BS transmitting for the whole time-window of 6 minutes, somehow modelling the downlink (DL) band of a frequency division duplex (FDD) system. In this kind of analysis, RF exposure in FDD is higher when compared to the exposure in a TDD setup, where DL and UL transmissions share the same band. Indeed, in TDD, the compliance distances turn out to be smaller simply because the BSs do not transmit on the time slots allocated to UL transmissions. Just as an example, if we consider a TDD setup with a DL to UL ratio of 0.75, i.e., with the BSs not transmitting 25\% of the time, the two compliance distances mentioned above (computed with the traditional method) will reduce to 13.9 m and 7.8 m in UMa and UMi, respectively. Moreover, with fully loaded BSs, the required compliance distance in TDD further decreases when the DL to UL ratio decreases. As a consequence, in the rest of the paper we consider BSs always transmitting for the whole time-window as this setup is characterized by the highest RF exposure.

\begin{figure}[t!]
\centering
\includegraphics[width=1\hsize]{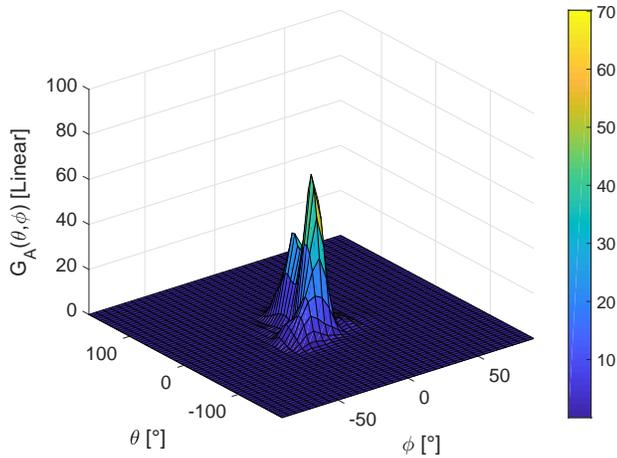}\\
\caption{Realization of $G_{\rm A}(\theta,\phi)$ in UMa when $K=1$ and $D=60$ s.}
\label{fig_rp_k1_d60}
\end{figure}

The radiation pattern of a BS in realistic operation looks very different from the one of Fig. \ref{fig_rp_art}. In Fig. \ref{fig_rp_k1_d60} we show a realization of the radiation pattern for a BS in UMa scenario serving in each time slot only $K=1$ UE with a drop duration $D=60$ s: in this case, the radiation pattern includes the combination of beamformers used to serve a sequence of 6 UEs that might be in very different channel conditions. The example of Fig. \ref{fig_rp_k1_d60} shows two main peaks, probably related to two UEs in different locations and both in line of sight (LOS) conditions. More importantly, the maximum average beamforming gain is about 70 in linear scale, which corresponds to about 18 dBi in logarithmic scale, and which is six times less when compared to the one obtained with the traditional method in Fig. \ref{fig_rp_art}.

\begin{figure}[t!]
\centering
\includegraphics[width=1\hsize]{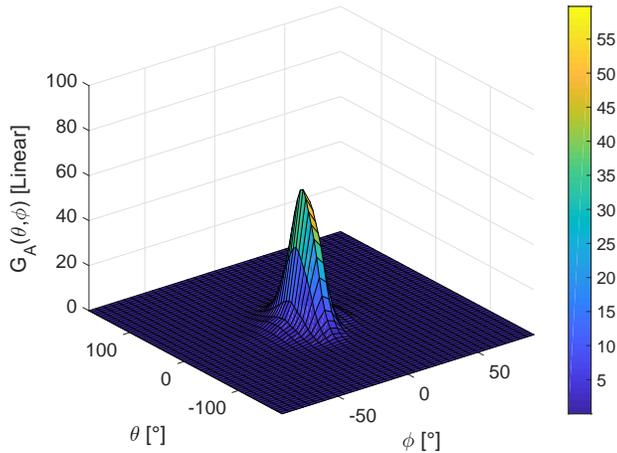}\\
\caption{Realization of $G_{\rm A}(\theta,\phi)$ in UMa when $K=5$ and $D=1$ s.}
\label{fig_rp_k5_d1}
\end{figure}

Then, in Fig. \ref{fig_rp_k5_d1} we show a realization of the radiation pattern of a BS still in UMa scenario, but serving now $K=5$ UEs in each time slot, with a drop duration $D=1$ s. Differently from Fig. \ref{fig_rp_k1_d60}, the number of UEs served over the time window is very high here, with the average including many possible UE locations and channel conditions. Indeed, the shape of the radiation pattern is quite regular, with a rather broad beam, in particular in the azimuth domain, which covers the whole BS sector, and a maximum average beamforming gain of about 60 in linear scale, again much smaller when compared to the traditional method.

To better understand how the compliance boundary depends on the parameters $K$ and $D$, we focus now on the direction $(\theta^{(\rm max)},\phi^{(\rm max)})$ that maximizes the power density, and, in turn, the compliance distance (\ref{eq_rez}). In both UMa and UMi scenarios, UEs are uniformly distributed in the azimuth domain and, because of the shape of the beamforming gain of each antenna element, the maximum power density is observed exactly at $\theta_{\rm UMa}^{(\rm max)}=\theta_{\rm UMi}^{(\rm max)}=0^{\circ}$. On the other hand, in the elevation domain, UEs are distributed either outdoor on the streets or inside buildings, whose maximum height is 24 m (computed by assuming a floor height of 3 m and a maximum of 8 floors per building) \cite[Tab. 6-1]{3gpp_tr36873_grangeat}. As in UMa BSs are deployed at a height of 25 m, the elevation angle of the BS-UE LOS link is always negative, and the maximum power density is observed at $\phi_{\rm UMa}^{(\rm max)}=-5^{\circ}$. On the contrary, in UMi BSs are deployed at a smaller height of 10 m, the elevation angle of the BS-UE LOS link can also be positive, and the maximum power density is observed at $\phi_{\rm UMi}^{(\rm max)}=0^{\circ}$.

In the following, in addition to the average beamforming gain $G_{\rm A}(\theta,\phi)$, we focus also on the normalized actual transmit power in the direction of maximum $P_{\rm A}(\theta^{(\rm max)},\phi^{(\rm max)})$, which is the metric mostly considered by RF exposure standards bodies like the IEC \cite{iec_fdis62232_new_grangeat}. This normalized power can be computed as
\begin{equation}
P_{\rm A}(\theta^{(\rm max)},\phi^{(\rm max)}) = \frac{G_{\rm A}(\theta^{(\rm max)},\phi^{(\rm max)})}{G_{\rm MAX}}\,,
\end{equation}
where $G_{\rm MAX}$ is the maximum beamforming gain that the BS can offer. This parameter depends on the antenna element gain and on the number of BS antennas, and, as shown in Fig. \ref{fig_rp_art}, is about 26 dBi in the considered system. 

\begin{figure}[t!]
\centering
\includegraphics[width=1\hsize]{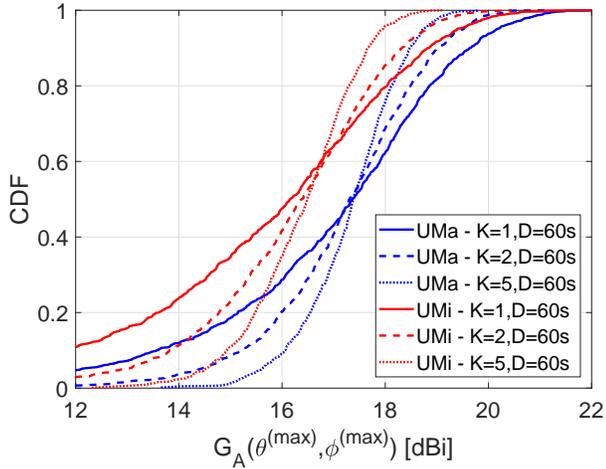}\\
\caption{CDF of $G_{\rm A}(\theta^{(\rm max)},\phi^{(\rm max)})$ for both UMa and UMi when $D=60$ s and for $K=1,2,5$.}
\label{fig_cdf_g_d60}
\end{figure}

In Fig. \ref{fig_cdf_g_d60} we report the CDF of the average beamforming gain $G_{\rm A}(\theta,\phi)$ in the direction of maximum power density $(\theta^{(\rm max)},\phi^{(\rm max)})$ for different values of $K$ and when $D=60$ s. First, we observe that the expected value of $G_{\rm A}(\theta^{(\rm max)},\phi^{(\rm max)})$ is far higher when compared to the antenna element gain of 8 dBi: in fact, the expected value would be equal to the antenna element gain only if the UEs served by the BSs were uniformly distributed not only in the azimuth, but also in the elevation domain. As the elevation angle of the BS-UE LOS link spans a range of just about $25^{\circ}$ and $40^{\circ}$ for most of the UEs \cite[Fig. 8.2-6]{3gpp_tr36873_grangeat} in UMa and UMi, respectively, BSs tend to focus the power mainly toward these specific elevation directions. Indeed, the median value of the UMa curves is about 17.4 dBi, whereas in UMi, because of the larger elevation angle range, the median value is about 1 dB less. Then, in Fig. \ref{fig_cdf_g_d60} we observe, as expected, that the slope of the curve increases, or, equivalently, the variance of the average beamforming gain decreases, when $K$ increases. For instance, when looking at the $p=99$th percentile of the UMa CDF curves, the average beamforming gain is about $21$ dBi with $K=1$ and about $19$ dBi with $K=5$.

\begin{figure}[t!]
\centering
\includegraphics[width=1\hsize]{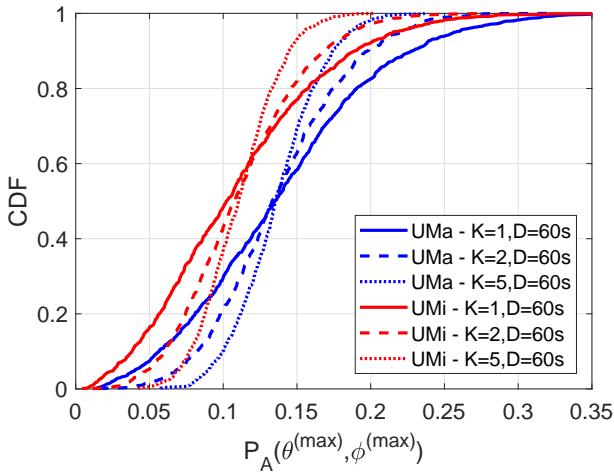}\\
\caption{CDF of $P_{\rm A}(\theta^{(\rm max)},\phi^{(\rm max)})$ for both UMa and UMi when $D=60$ s and for $K=1,2,5$.}
\label{fig_cdf_p_d60}
\end{figure}

In Fig. \ref{fig_cdf_p_d60} we show the CDF of the normalized actual transmit power $P_{\rm A}(\theta^{(\rm max)},\phi^{(\rm max)})$ when $D=60$ s. By looking at the case with $K=1$, this figure shows that the 95th percentile of the actual transmit power is about 26\% and 22\%  of the maximum transmit power in UMa and UMi, respectively. These percentages increase respectively to 32\% and 27\% when the 99th percentile is considered.

\begin{figure}[t!]
\centering
\includegraphics[width=1\hsize]{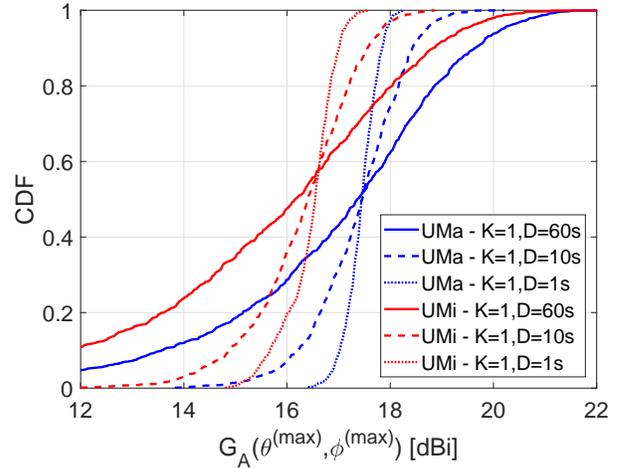}\\
\caption{CDF of $G_{\rm A}(\theta^{(\rm max)},\phi^{(\rm max)})$ for both UMa and UMi when $K=1$ and for $D=1,10,60$ s.}
\label{fig_cdf_g_k1}
\end{figure}

\begin{figure}[t!]
\centering
\includegraphics[width=1\hsize]{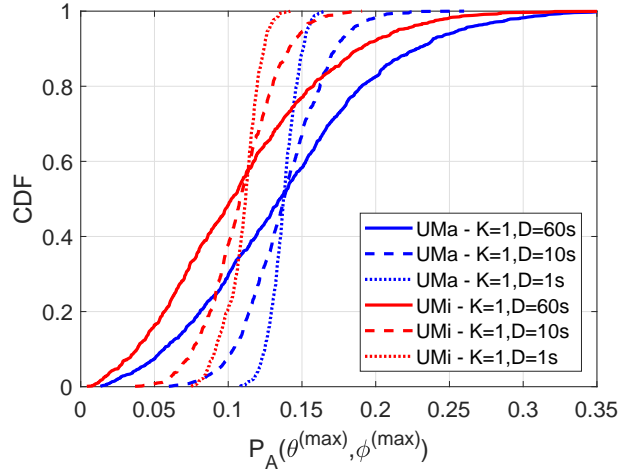}\\
\caption{CDF of $P_{\rm A}(\theta^{(\rm max)},\phi^{(\rm max)})$ for both UMa and UMi when $K=1$ and for $D=1,10,60$ s.}
\label{fig_cdf_p_k1}
\end{figure}

In Fig.s \ref{fig_cdf_g_k1} and \ref{fig_cdf_p_k1} we set $K=1$ and show, for different values of $D$, the CDF of $G_{\rm A}(\theta^{(\rm max)},\phi^{(\rm max)})$ and $P_{\rm A}(\theta^{(\rm max)},\phi^{(\rm max)})$, respectively. Reducing the drop duration allows to further decrease the variance of the average beamforming gain: for instance, when still looking at the $p=99$th percentile of the UMa curves, the average beamforming gain is about $18$ dBi with $D=1$ s.

Finally, in Tab.s \ref{tab_ez} and \ref{tab_ez_perc} we report the compliance distance (\ref{eq_rez}) for different traffic conditions computed by using the proposed statistical approach both as an absolute value and as a percentage of the distance obtained with the traditional conservative method. First, we note that smaller compliance boundaries are required in UMi when compared to UMa because of the lower BS transmit power: 44 dBm compared to 49 dBm. Then, smaller compliance boundaries are required also when $K$ increases and when $D$ decreases. However, even with just $K=1$ UE active per BS and a drop duration $D=60$ s, the compliance distance when looking at the $p=99$th percentile is about 56\% and 52\% of the distance required by the traditional method in UMa and UMi, respectively.

These results show that statistical approaches, by taking into account actual BS operations, allow to strongly reduce the dimension of the compliance boundary around massive MIMO BSs, thus facilitating their deployment in the sites currently used by GSM, UMTS or LTE.

\begin{table}[t]
\centering
\caption{Summary of the compliance distance $r_{{\rm CB},p}$ in UMa and UMi scenarios.}
\begin{tabular}{|l|c|c|c|c|}
\hline
                & \multicolumn{4}{c|}{$r_{{\rm CB},p}$ [m]} \\ \hline
                & \multicolumn{2}{c|}{UMa} & \multicolumn{2}{c|}{UMi} \\ \hline
Configuration   & $p=95$th 	& $p=99$th 	& $p=95$th 	& $p=99$th	\\ \hline
$K=1$, $D=1$ s  & 6.3 		& 6.4 		& 3.2 		& 3.3		\\ \hline
$K=1$, $D=10$ s & 6.9 		& 7.3 		& 3.5 		& 3.8		\\ \hline
$K=1$, $D=60$ s & 8.1 		& 9.0 		& 4.2 		& 4.7		\\ \hline
$K=2$, $D=1$ s  & 6.2 		& 6.3 		& 3.2 		& 3.2		\\ \hline
$K=2$, $D=10$ s & 6.6 		& 6.9 		& 3.4 		& 3.6		\\ \hline
$K=2$, $D=60$ s & 7.6 		& 8.0 		& 3.9 		& 4.3		\\ \hline
$K=5$, $D=1$ s  & 6.1 		& 6.2 		& 3.1 		& 3.1		\\ \hline
$K=5$, $D=10$ s & 6.4 		& 6.6 		& 3.3 		& 3.4		\\ \hline
$K=5$, $D=60$ s & 6.9 		& 7.3 		& 3.6 		& 3.8		\\ \hline
\end{tabular}
\label{tab_ez}
\end{table}

\begin{table}[t]
\centering
\caption{Summary of the compliance distance $r_{{\rm CB},p}$ as a percentage of the one obtained with the traditional conservative method. Note that $r_{{\rm CB,UMa}}^{\rm (trad)}=16.1$ m and $r_{{\rm CB,UMi}}^{\rm (trad)}=9$ m.}
\begin{tabular}{|l|c|c|c|c|}
\hline
                & \multicolumn{4}{c|}{$r_{{\rm CB},p}/r_{{\rm CB}}^{\rm (trad)}$ [\%]} \\ \hline
                & \multicolumn{2}{c|}{UMa} & \multicolumn{2}{c|}{UMi} \\ \hline
Configuration   & $p=95$th 	& $p=99$th 	& $p=95$th 	& $p=99$th	\\ \hline
$K=1$, $D=1$ s  & 39 		& 40 		& 36 		& 37		\\ \hline
$K=1$, $D=10$ s & 43 		& 45 		& 39 		& 42		\\ \hline
$K=1$, $D=60$ s & 50 		& 56 		& 47 		& 52		\\ \hline
$K=2$, $D=1$ s  & 39 		& 39 		& 36 		& 36		\\ \hline
$K=2$, $D=10$ s & 41 		& 43 		& 38 		& 40		\\ \hline
$K=2$, $D=60$ s & 47 		& 50 		& 43 		& 48		\\ \hline
$K=5$, $D=1$ s  & 38 		& 39 		& 34 		& 34		\\ \hline
$K=5$, $D=10$ s & 40 		& 41 		& 37 		& 38		\\ \hline
$K=5$, $D=60$ s & 43 		& 45 		& 40 		& 42		\\ \hline
\end{tabular}
\label{tab_ez_perc}
\end{table}

\section{Conclusions}
In this paper, we have proposed a statistical approach for assessing the RF exposure conditions around massive MIMO BSs based on the 3D spatial channel model developed by 3GPP. The methodology consists in performing system simulations that take into account realistic deployment scenarios in terms of installation height, UE distribution and traffic, to evaluate the CDF of the BS actual transmit power. The compliance boundary is then computed for a given percentile of the resulting BS transmit power. By considering only one active UE per BS, numerical results show that the 95th/99th percentile of the actual BS transmit power is just 26\%/32\% in UMa and 22\%/27\% in UMi of the maximum transmit power, and that allows reducing to nearly half the compliance distance. Moreover, the dimension of the compliance boundary further decreases when a higher number of active UEs is served by the BS or a smaller drop duration is examined. The proposed statistical approach contributes to improve the calculation methods already defined in \cite{iec_fdis62232_new_grangeat} and support the deployment of massive MIMO BSs for 5G and beyond cellular networks. As a concluding remark, we highlight that all the statistical approaches including ours, although based on realistic assumptions, require anyhow complementary techniques, based for instance on power control and beamforming adaptation \cite[Sect. V]{sambo_st15}, to ensure that the EMF constraints are met at the BSs for all the possible actual configurations.


\bibliographystyle{IEEEtran}
\bibliography{IEEEabrv,full_bibliography}

\begin{thebibliography}{10}
\providecommand{\url}[1]{#1}
\csname url@samestyle\endcsname
\providecommand{\newblock}{\relax}
\providecommand{\bibinfo}[2]{#2}
\providecommand{\BIBentrySTDinterwordspacing}{\spaceskip=0pt\relax}
\providecommand{\BIBentryALTinterwordstretchfactor}{4}
\providecommand{\BIBentryALTinterwordspacing}{\spaceskip=\fontdimen2\font plus
\BIBentryALTinterwordstretchfactor\fontdimen3\font minus
  \fontdimen4\font\relax}
\providecommand{\BIBforeignlanguage}[2]{{%
\expandafter\ifx\csname l@#1\endcsname\relax
\typeout{** WARNING: IEEEtran.bst: No hyphenation pattern has been}%
\typeout{** loaded for the language `#1'. Using the pattern for}%
\typeout{** the default language instead.}%
\else
\language=\csname l@#1\endcsname
\fi
#2}}
\providecommand{\BIBdecl}{\relax}
\BIBdecl

\bibitem{larsson_cm14}
E.~G. Larsson, O.~Edfors, F.~Tufvesson, and T.~L. Marzetta, ``Massive {MIMO}
  for next generation wireless systems,'' \emph{{IEEE} Commun. Mag.}, vol.~52,
  no.~2, pp. 186--195, Feb. 2014.

\bibitem{nam_cm12}
Y.~H. Nam, B.~L. Ng, K.~Sayana, Y.~Li, J.~Zhang, Y.~Kim, and J.~Lee,
  ``Full-dimension {MIMO} ({FD-MIMO}) for next generation cellular
  technology,'' \emph{{IEEE} Commun. Mag.}, vol.~51, no.~6, pp. 172--179, Jun.
  2013.

\bibitem{sambo_st15}
Y.~A. Sambo, F.~Héliot, and M.~A. Imran, ``A survey and tutorial of
  electromagnetic radiation and reduction in mobile communication systems,''
  \emph{{IEEE} Commun. Surveys Tuts.}, vol.~17, no.~2, pp. 790--802,
  Secondquarter 2015.

\bibitem{emf_2b}
``{IARC} classifies radiofrequency electromagnetic fields as possibly
  carcinogenic to humans,'' Press Release No. 208, May 2011.

\bibitem{vallauri_map15}
R.~Vallauri, G.~Bertin, B.~Piovano, and P.~Gianola, ``Electromagnetic field
  zones around an antenna for human exposure assessment: evaluation of the
  human exposure to {EMFs},'' \emph{{IEEE} Antennas Propag. Mag.}, vol.~57,
  no.~5, pp. 53--63, Oct. 2015.

\bibitem{iec_fdis62232_new_grangeat}
{IEC 62232:2017}, ``{Determination of RF field strength, power density and SAR
  in the vicinity of radiocommunication base stations for the purpose of
  evaluating human exposure},'' Aug. 2017.

\bibitem{icnirp_guidelines_grangeat}
{International Commission on Non-Ionizing Radiation Protection}, ``Guidelines
  for limiting exposure to time-varying electric, magnetic, and electromagnetic
  fields (up to 300 {GHz}),'' \emph{Health Physics}, vol.~74, no.~4, pp.
  494--522, Apr. 1998.

\bibitem{blanch_tap12}
S.~Blanch, J.~Romeu, and A.~Cardama, ``Near field in the vicinity of wireless
  base-station antennas: an exposure compliance approach,'' \emph{{IEEE} Trans.
  Antennas Propag.}, vol.~50, no.~5, pp. 685--692, May 2002.

\bibitem{degirmenci_tec16}
E.~Degirmenci, B.~Thors, and C.~T{\"{o}}rnevik, ``Assessment of compliance with
  {RF} {EMF} exposure limits: approximate methods for radio base station
  products utilizing array antennas with beam-forming capabilities,''
  \emph{{IEEE} Trans. Electromagn. Compat.}, vol.~58, no.~4, pp. 1110--1117,
  Aug. 2016.

\bibitem{perentos_tap12}
N.~Perentos, S.~Iskra, A.~Faraone, R.~J. McKenzie, G.~Bit-Babik, and
  V.~Anderson, ``Exposure compliance methodologies for multiple input multiple
  output ({MIMO}) enabled networks and terminals,'' \emph{{IEEE} Trans.
  Antennas Propag.}, vol.~60, no.~2, pp. 644--653, Feb. 2012.

\bibitem{thors_bio14}
B.~Thors, A.~Thielens, J.~Frid{\'{e}}n, D.~Colombi, C.~T{\"{o}}rnevik,
  G.~Vermeeren, L.~Martens, and W.~Joseph, ``Radio frequency electromagnetic
  field compliance assessment of multi-band and {MIMO} equipped radio base
  stations,'' \emph{Bioelectromagnetics}, vol.~35, no.~4, pp. 296--308, May
  2014.

\bibitem{thors_acc17}
B.~Thors, A.~Furusk{\"{a}}r, D.~Colombi, and C.~T{\"{o}}rnevik, ``Time-averaged
  realistic maximum power levels for the assessment of radio frequency exposure
  for {5G} radio base stations using massive {MIMO},'' \emph{IEEE Access},
  vol.~5, pp. 19\,711--19\,719, 2017.

\bibitem{altman_tec02}
Z.~Altman, B.~Begasse, C.~Dale, A.~Karwowski, J.~Wiart, M.-F. Wong, and
  L.~Gattoufi, ``Efficient models for base station antennas for human exposure
  assessment,'' \emph{{IEEE} Trans. Electromagn. Compat.}, vol.~44, no.~4, pp.
  588--592, Nov. 2002.

\bibitem{3gpp_tr36873_grangeat}
{3GPP TR 36.873}, ``{Study on 3D channel model for LTE (Release 12)},'' Jun.
  2015.

\bibitem{zhou_it03}
S.~Zhou and G.~B. Giannakis, ``Optimal transmitter eigen-beamforming and
  space-time block coding based on channel correlations,'' \emph{{IEEE} Trans.
  Inf. Theory}, vol.~49, no.~7, pp. 1673--1690, Jul. 2003.

\end{thebibliography}
		
\end{document}